\documentclass[prd,twocolumn,showpacs,footinbib]{revtex4}
\usepackage{amsmath}
\usepackage{graphicx}

\newcommand\gr[1]{\mathrm{#1}} % notation for groups
\newcommand\pV{\mathcal{V}} % physical effective potential
\DeclareMathOperator{\tr}{Tr}
\DeclareMathOperator{\li}{li}
\newcommand\La{\mathcal L}
\newcommand\de{\partial}
\def\sumint{\hbox{$\sum$}\!\!\!\!\!\!\int}
\begin{document}

\title{Linear sigma model at finite density in the $1/N$ expansion to
next-to-leading order}
\author{Jens O. Andersen}
\email{andersen@tf.phys.ntnu.no}
\affiliation{Department of Physics, Norwegian University of Science and Technology, N-7491
Trondheim, Norway}
\author{Tom\'{a}\v{s} Brauner}
\email{brauner@ujf.cas.cz}
\affiliation{Institut f\"ur Theoretische Physik, J. W. Goethe-Universit\"at, Max von
Laue-Stra\ss e 1, D-60438 Frankfurt am Main, Germany\footnote{On leave from Department of
Theoretical Physics, Nuclear Physics Institute ASCR, 25068 \v Re\v z, Czech Republic}}

\begin{abstract}
We study relativistic Bose--Einstein condensation at finite density and
temperature using the linear sigma model in the one-particle-irreducible
$1/N$-expansion. We derive the effective potential to next-to-leading (NLO)
order and show that it can be renormalized in a temperature-independent manner.
As a particular application, we study the thermodynamics of the pion gas in the
chiral limit as well as with explicit symmetry breaking. At nonzero temperature
we solve the NLO gap equation and show that the results describe the
chiral-symmetry-restoring second-order phase transition in agreement with
general universality arguments. However, due to nontrivial regularization
issues, we are not able to extend the NLO analysis to nonzero chemical
potential.
\end{abstract}

\maketitle

\section{Introduction}
The Lagrangian of quantum chromodynamics (QCD) with $N_f$ quark flavors has an
extended $\gr{SU}(N_f)_L\times\gr{SU}(N_f)_R$ symmetry in the chiral limit,
which is not respected by the true QCD ground state. The chiral symmetry is
spontaneously broken down to the vector subgroup $\gr{SU}(N_f)_V$ by quantum
effects, and according to Goldstone's theorem there are $N_f^2-1$ massless
excitations associated with this breaking. For $N_f=2$, this implies the
existence of three Goldstone bosons, which are identified with the pions. The
fact that these particles are not exactly massless is taken into account by
small nonzero quark masses in the QCD Lagrangian. In low-energy QCD
phenomenology, one takes advantage of the isomorphism between the groups
$\gr{SU}(2)\times\gr{SU}(2)$ and $\gr{SO}(4)$ to describe the pions as well as
the sigma by scalar field theory with this symmetry. At low temperature, the
$\gr{O}(4)$ symmetry is broken down to $\gr{O}(3)$ either explicitly or
spontaneously. In the latter case, one expects it to be restored at
sufficiently high temperature. Lattice simulations of QCD suggest that the
critical temperature is around $170\text{ MeV}$, depending on the masses of the
quarks.

At zero temperature, the $\gr{O}(N)$ linear sigma model was studied a long time
ago using the one-particle-irreducible (1PI) $1/N$-expansion, to leading order
(LO) by Coleman {\it et al.}~\cite{Coleman:1974jh}, and to next-to-leading
order (NLO) by Root~\cite{Root:1974zr}. At finite temperature but vanishing
chemical potential, the LO thermodynamics was studied by Meyer-Ortmanns {\it et
al.}~\cite{MeyersOrtmanns:1993dw}. Later several attempts to calculate the
pressure to next-to-leading order have been made, but the approximations made
were uncontrolled and renormalization was ignored. Due to the nontrivial
renormalization issues, the NLO pressure was calculated only
recently~\cite{Andersen:2004ae}. Specifically, one problem was that the
effective potential seemed to have ultraviolet divergences which were
temperature-dependent, except at its minimum. Renormalization of the effective
potential could therefore be carried out with temperature-independent
counterterms only on shell.

Sigma models have also been investigated in detail at finite temperature using
other methods: Optimized perturbation theory~\cite{Chiku:1998kd},
two-particle-irreducible (2PI) effective action either in the Hartree or
large-$N$ approximation
\cite{Cornwall:1974vz,AmelinoCamelia:1992nc,AmelinoCamelia:1997dd,Petropoulos:1998gt,
Lenaghan:1999si}. In these calculations, one typically encounters a local gap
equation for a mass parameter, which is straightforward to solve. Beyond
leading order, the resulting equations are nonlocal and much harder to solve.

Linear sigma models at finite chemical potential that describe relativistic
Bose--Einstein condensation, have received interest in recent years due to the
possibility of e.g. pion and kaon condensation in compact stars
\cite{Kaplan:2001qk,Bedaque:2001je}. Pion condensation and the phase diagram of
two-flavor QCD have been investigated using chiral perturbation theory~
\cite{Son:2000xc,Splittorff:2000mm,Kogut:2001id,Loewe:2002tw,Loewe:2004mu,Loewe:2005yn},
lattice QCD~\cite{Kogut:2002zg,Gupta:2002kp}, ladder
QCD~\cite{Barducci:2003un}, chiral quark model~\cite{Jakovac:2003ar},
Nambu--Jona-Lasinio models in the mean-field
approximation~\cite{Nambu:1961tp,Hufner:1994ma,Zhuang:1994dw,Barducci:2004tt,He:2005nk,
Ebert:2005cs,Ebert:2005wr,Andersen:2007qv}, and the linear sigma model using
the Hartree approximation~\cite{Mao:2006zr}, or the large-$N$
approximation~\cite{Andersen:2006ys}.

In the present paper, we extend the calculation of Ref. \cite{Andersen:2004ae}
using the 1PI-$1/N$ expansion by including nonzero chemical potential. In
contrast to the previous work we show that, when interpreted properly, the
effective potential can be renormalized in a temperature-independent manner
even off its minimum. We apply the method to the study of pion condensation.
The paper is organized as follows. In Sec.~\ref{Sec:eff_act}, we introduce the
general model and briefly discuss interacting bosons at finite temperature and
density. In Sec.~\ref{Sec:LO}, we review the standard calculation of the LO
effective potential and determine the phase diagram. In Sec.~\ref{Sec:NLO}, we
calculate the NLO effective potential and discuss in detail its
renormalization. In particular, we demonstrate how it is used to derive a
renormalized gap equation. In Sec.~\ref{Sec:Concl} we summarize and conclude.

\section{Effective action in $1/N$ expansion}
\label{Sec:eff_act}

\subsection{The model}
\label{Subsec:model}
The Euclidean Lagrangian for a Bose gas of $N$ real scalars is given by
\cite{Andersen:2004ae}
\begin{equation}
\La=\frac12(\de_\mu\phi_i)^2+\frac{\lambda_b}{8N}(\phi_i\phi_i-Nf_{\pi,b}^2)^2
-\sqrt{N}H\phi_N,
\label{l1}
\end{equation}
where $i=1,\dotsc,N$. The parameters $f_{\pi,b}$ and $\lambda_b$ denote the
bare pion decay constant and coupling, respectively. If $H=0$, the Lagrangian
has an $\gr{O}(N)$ symmetry, while if $H\neq0$, this symmetry is explicitly
broken down to $\gr{O}(N-1)$. Each generator of the symmetry group gives rise
to a conserved charge and for $\gr{O}(N)$ there are $N(N-1)/2$ such charges. We
can characterize a system described by Eq.~(\ref{l1}) by the expectation values
of the conserved charges $Q_i$ and for each of them we can in principle
introduce a chemical potential $\mu_i$. Nevertheless, it is only possible to
simultaneously specify two different charges if they commute. For the
$\gr{O}(N)$ symmetry the maximum number of commuting charges is $\lfloor
N/2\rfloor$ (here $\lfloor\,\rfloor$ denotes the integer part of a real
number)~\cite{Haber:1981ts}.

If we treat two real scalar fields, $\phi_R$ and $\phi_I$, as the real and
imaginary part of a single complex field, $\Phi=(\phi_R+i\phi_I)/\sqrt2$, the
conserved charge $Q$ can be incorporated by making the substitution in the
Euclidean Lagrangian,
\begin{equation}
\begin{split}
\de_0\Phi&\to(\de_0-\mu)\Phi,\\
\de_0\Phi^\dagger&\to(\de_0+\mu)\Phi^\dagger.
\end{split}
\label{musub}
\end{equation}
The actual value of the charge density in thermodynamic equilibrium is then
determined from the free energy density ${\cal F}$ as $Q=-{\de{\cal
F}}/{\de\mu}$.

In order to eliminate the quartic term from the Lagrangian (\ref{l1}), we
introduce an auxiliary field $\alpha$ and add to Eq.~(\ref{l1}) the following
term~\cite{Coleman:1974jh},
\begin{equation*}
\La_\alpha=\frac N{2\lambda_b}\left[\alpha-\frac{i\lambda_b}{2N}(\phi_i\phi_i
-Nf^2_{\pi,b})\right]^2.
\end{equation*}
The Lagrangian can now be written as
\begin{equation}
\La=\frac12(\de_\mu\phi_i)^2-\frac i2\alpha(\phi_i\phi_i-Nf_{\pi,b}^2)+\frac
N{2\lambda_b}\alpha^2-\sqrt NH\phi_N.
\label{auxLa}
\end{equation}
By using the equation of motion for $\alpha$, one can eliminate this field
altogether and thus recover the original Lagrangian~(\ref{l1}).

Another virtue of introducing the auxiliary field is that the Lagrangian now
depends linearly on the two parameters, $f^2_{\pi,b}$ and $1/\lambda_b$. As
will be shown later, the divergences that appear in the effective potential may
be removed by their suitable redefinition. For the sake of systematic expansion
in powers of $1/N$, we write them as
\begin{equation}
\begin{split}
f_{\pi,b}^2&=f_{\pi}^2+a_0+\frac1N a_1+\dotsb,\\
\frac1{\lambda_b}&=\frac1\lambda+b_0+\frac1N b_1+\dotsb,
\end{split}
\label{counterterms}
\end{equation}
where $f_\pi$ and $\lambda$ are the \emph{renormalized} decay constant and
coupling. The parameters $a_{0,1}$ and $b_{0,1}$ are the counterterms needed to
cancel the divergences at the corresponding orders in $1/N$.

\subsection{Application to pion gas}
\label{Subsec:pigas} We next use our model to describe the pion gas at finite
temperature and chemical potential. The vacuum expectation value $\phi_0$ of
$\phi_N$ is interpreted as the usual chiral condensate. Using Eq. (\ref{musub})
we introduce isospin chemical potential $\mu_I$ for the pair $\phi_1,\phi_2$,
which are hence identified with the real and imaginary parts of the charged
pion field. The neutral pion is $\phi_3$. The Lagrangian (\ref{auxLa}) thus
becomes
\begin{multline*}
\La=\frac12(\de_\mu\phi_i)^2-\frac i2\alpha(\phi_i\phi_i-Nf_{\pi,b}^2)+\frac
N{2\lambda_b}\alpha^2-\sqrt NH\phi_N\\
-i\mu_I(\phi_1\de_0\phi_2-\phi_2\de_0\phi_1)-\frac12\mu_I^2(\phi_1^2+\phi_2^2).
\end{multline*}
The Lagrangian is quadratic in the scalar fields $\phi_i$ and we next integrate
over the $N-3$ fields $\phi_3,\dotsc,\phi_{N-1}$. This gives an effective
action for the remaining fields $\alpha,\phi_1,\phi_2,\phi_N$,
\begin{multline}
S_{\text{eff}}=\frac12(N-3)\tr\log(-\de^2-i\alpha)+\int_0^\beta
d\tau\int d^3x\\
\times\left[\frac12(\de_\mu\phi_1)^2+\frac12(\de_\mu\phi_2)^2
+\frac12(\de_\mu\phi_N)^2-\sqrt{N}H\phi_N\right.\\
-i\mu_I(\phi_1\de_0\phi_2-\phi_2\de_0\phi_1)-\frac12\mu_I^2(\phi_1^2+\phi_2^2)\\
\left.+\frac
N{2\lambda_b}\alpha^2-\frac i2\alpha(\phi_1^2+\phi_2^2+\phi_N^2-Nf_{\pi,b}^2)\right].
\label{efa}
\end{multline}
Now we introduce a nonzero expectation value $\rho_0$ for $\phi_1$ to allow for
a nonzero pion condensate. We also expand the auxiliary field $\alpha$ around
its expectation value $iM^2$. Thus we make the following replacements in Eq.
(\ref{efa}),
\begin{equation}
\begin{split}
\phi_1&\to\sqrt{N}\rho_0+\phi_1,\\
\phi_N&\to\sqrt{N}\phi_0+\phi_N,\\
\alpha&\to iM^2+\frac\alpha{\sqrt{N}}.
\end{split}
\label{subs}
\end{equation}
The various factors of $\sqrt{N}$ do not change the physics, but merely
facilitate the calculations in powers of $1/N$. Making the substitutions
(\ref{subs}) in Eq.~(\ref{efa}), the effective action becomes~\footnote{We are
omitting terms linear in the quantum fields as these terms vanish at the
minimum of the effective potential.}
\begin{widetext}
\begin{multline}
S_{\text{eff}}=\frac12(N-3)\tr\log\left(-\de^2+M^2-\frac{i\alpha}{\sqrt{N}}\right)
+\int_0^\beta d\tau\int d^3x\left[\frac12(\de_\mu\phi_1)^2
+\frac12(\de_{\mu}\phi_2)^2+\frac12(\partial_{\mu}\phi_N)^2\right.\\
-NH\phi_0-i\mu_I(\phi_1\de_0\phi_2-\phi_2\de_0\phi_1)-
\frac12\mu_I^2(N\rho_0^2+\phi_1^2+\phi_2^2)
-\frac N{2\lambda_b}M^4+\frac1{2\lambda_b}\alpha^2
+\frac12M^2(\phi_1^2+\phi_2^2+\phi_N^2)\\
\left.-i\alpha\left(\phi_0\phi_N+\rho_0\phi_1+
\frac{\phi_1^2+\phi_2^2+\phi_N^2}{2\sqrt{N}}\right)
+\frac12NM^2(\phi_0^2+\rho_0^2-f_{\pi,b}^2)\right].
\label{ea}
\end{multline}
Expanding Eq.~(\ref{ea}) up to second order in $1/\sqrt{N}$ and using Eq.
(\ref{counterterms}), we obtain
\begin{multline}
\frac{S_{\text{eff}}}{\beta V}=\frac12NM^2(\phi_0^2+\rho_0^2-f_{\pi}^2)
-\frac{NM^4}{2\lambda}-NH\phi_0-\frac12N\mu_I^2\rho_0^2+
\frac12(N-3)\sumint_P\ln(P^2+M^2)-\frac12NM^2a_0-\frac12NM^4b_0\\
+\frac12\sumint_P\chi^T(-P)\left(
\begin{array}{cccc}
\frac12\Pi(P,M)+\frac{1}{\lambda}+b_0 & -i\phi_0 & -i\rho_0 & 0\\
-i\phi_0 & P^2+M^2 & 0 & 0\\
-i\rho_0 & 0 & P^2+m^2 & -2\mu_I\omega_n\\
0 & 0 & 2\mu_I\omega_n & P^2+m^2\\
\end{array}\right)\chi(P)-\frac12M^2a_1-\frac12M^4b_1,
\label{matrix}
\end{multline}
\end{widetext}
where $\chi(P)=(\alpha(P),\phi_N(P),\phi_1(P),\phi_2(P))^T$ and $m^2=M^2-\mu_I^2$. The
self-energy function is
\begin{equation*}
\Pi(P,M)=\sumint_Q\frac1{Q^2+M^2}\frac1{(P+Q)^2+M^2}.
\end{equation*}
We have also introduced the notation
\begin{equation*}
\sumint_P=T\sum_{\omega_n=2\pi nT}\int\frac{d^3p}{(2\pi)^3}.
\end{equation*}
The sum-integral involves a summation over Matsubara frequencies and an
integral over three-momenta $p$. At zero temperature, the sum-integrals reduce
to a four-dimensional integral, where we write
\begin{equation*}
\int_P=\int\frac{d^4P}{(2\pi)^4}.
\end{equation*}
Such integrals are regularized by a four-dimensional ultraviolet cutoff
$\Lambda$.

In deriving Eq. (\ref{matrix}) we implicitly assumed that the classical fields
$\phi_0,\rho_0,M^2$ are constant. This makes it possible to define the
effective potential by dividing by the space volume and inverse temperature.
Specifically, we define its LO and NLO parts by
\begin{equation*}
-\frac{S_{\text{eff}}}{\beta V}=NV_{\text{LO}}+V_{\text{NLO}}.
\end{equation*}
In equilibrium, the effective potential is then equal to the pressure.

At this point it is worthwhile to discuss explicitly the symmetries of the
model with(out) the $H$ term and the chemical potential, and their spontaneous
breaking by the condensates $\phi_0,\rho_0$, in the physical case $N=4$. First,
in the chiral limit, $H=0$, the system has $\gr{O(4)}$ symmetry in the vacuum
which is broken explicitly by the chemical potential down to $\gr{O(2)\times
O(2)}$. The pion condensate (which is always favored in the chiral limit, see
Sec. \ref{Subsec:LO_phdiag}) breaks this down to $\gr{O(2)}$. The Goldstone
theorem then predicts a single exactly massless Goldstone boson, which is
identified with the charged pion (its antiparticle being massive). Only at zero
chemical potential is the unbroken symmetry larger, $\gr{O(3)}$, leading to
three Goldstone bosons.

Second, nonzero $H$ breaks the symmetry of the action explicitly down to
$\gr{O(3)}$. The chiral condensate is a singlet of this unbroken symmetry, and
therefore does not affect the low-energy spectrum. Nonzero isospin chemical
potential breaks the $\gr{O(3)}$ explicitly down to $\gr{O(2)}$, which is
subsequently broken spontaneously by the pion condensate, so we again find a
single massless Goldstone boson, to be identified with the charged pion.

Let us remark that, as is clear from the second line of Eq. (\ref{matrix}), the
above discussion of the excitation spectrum applies only after the NLO
correction to the effective action, and thus the contribution of the chemical
potential to the kinetic term, has been included. It can then be shown using
the NLO gap equation that the expected Goldstone bosons are indeed exactly
massless \cite{Andersen:2004ae}.

\section{Leading order}
\label{Sec:LO}

\subsection{Effective potential and gap equations}
\label{Subsec:LO_effpot} The LO part of the effective potential is given by the
first line of Eq. (\ref{matrix}),
\begin{multline*}
V_{\text{LO}}=\frac12M^2(f_{\pi}^2-\phi_0^2-\rho_0^2)
+\frac{M^4}{2\lambda}+H\phi_0+\frac12\mu_I^2\rho_0^2\\
-\frac12\sumint_P\ln(P^2+M^2)+\frac12M^2a_0+\frac12M^4b_0.
\end{multline*}
The sum-integral which appears here is evaluated as follows. First we extract
its zero-temperature part, i.e., the four-dimensional integral, which is
regulated using a four-dimensional cutoff. The difference of the integral over
$p_0$ and the Matsubara sum is UV-finite and calculated e.g. using standard
contour techniques. The result is
\begin{multline}
\sumint_P\ln(P^2+M^2)=\frac{M^2\Lambda^2}{16\pi^2}-\frac{M^4}{64\pi^2}
\left(2\ln\frac{\Lambda^2}{M^2}+1\right)\\
-\frac{T^4}{32\pi^2}J_0(\beta M),
\label{sumintlog}
\end{multline}
where we have dropped a (quartically divergent) constant piece as well as terms
suppressed by inverse powers of the cutoff, and defined
\begin{equation*}
J_0(\beta M)=\frac{32}{3T^4}\int_0^{\infty}dp\,\frac{p^4}{\omega_p}n(\omega_p).
\end{equation*}
Here $\omega_p=\sqrt{p^2+M^2}$, and $n(x)=1/(e^{\beta x}-1)$ is the Bose--Einstein
distribution function.

The divergences encountered in this calculation can be cancelled by adjusting
the LO counterterms $a_0,b_0$. We employ the ``minimal subtraction'' strategy,
i.e., just cancel the divergences in the effective potential, leaving the
finite part unchanged. This leads to
\begin{equation*}
a_0=\frac{\Lambda^2}{16\pi^2},\quad
b_0=-\frac1{32\pi^2}\ln\frac{\Lambda^2}{\mu^2},
\end{equation*}
where $\mu$ is the scale at which the coupling is renormalized. The
renormalized LO effective potential can then be written as
\begin{multline}
V_{\text{LO}}=\frac12M^2(f_{\pi}^2-\phi_0^2-\rho_0^2)
+\frac{T^4}{64\pi^2}J_0(\beta M)+H\phi_0\\
+\frac{M^4}{64\pi^2}\left(\frac{32\pi^2}\lambda+\ln\frac{\mu^2}{M^2}+\frac12\right)
+\frac12\mu_I^2\rho_0^2.
\label{VLO}
\end{multline}

Once we have renormalized the effective potential, the gap equations are
obtained by setting its derivatives with respect to the variables
$\phi_0,\rho_0$, and $M^2$ equal to zero,
\begin{align}
\label{dynamic_gapeq}
\frac{\de V}{\de\phi_0}&=0,\quad
\frac{\de V}{\de\rho_0}=0,\\
\label{constraint_gapeq}
\frac{\de V}{\de M^2}&=0.
\end{align}
As will be explained in detail in Sec. \ref{Subsec:NLO_ren}, it is Eq.
(\ref{dynamic_gapeq}) that represents the true dynamic gap equations, while Eq.
(\ref{constraint_gapeq}) can be treated as merely a constraint which serves to
eliminate the vacuum expectation value of the auxiliary field $\alpha$ in favor
of the condensates $\phi_0,\rho_0$. Using now the explicit expression for the
LO effective potential (\ref{VLO}), we find
\begin{align}
\label{renorm_gapeq_dynamic}
\phi_0M^2=H,&\quad \rho_0(M^2-\mu^2_I)=0,\\
\notag
f_{\pi}^2-\phi_0^2-\rho_0^2&-\frac{T^2}{16\pi^2}J_1(\beta M)\\
\label{renorm_gapeq_constraint}
&+\frac{M^2}{16\pi^2}\left(\frac{32\pi^2}\lambda+\ln\frac{\mu^2}{M^2}\right)=0,
\end{align}
where
\begin{equation*}
J_1(\beta M)=-\frac{T^2}2\frac{dJ_0(\beta M)}{dM^2}
=\frac8{T^2}\int_0^{\infty}dp\,\frac{p^2}{\omega_p}n(\omega_p).
\end{equation*}

\subsection{Phase diagram}
\label{Subsec:LO_phdiag}
\subsubsection{Chiral limit}
In the chiral limit and at zero chemical potential $\mu_I$, the Lagrangian is
exactly $\gr{O}(4)$-invariant and the $\phi_0$ and $\rho_0$ condensates
connected by a symmetry transformation. Since switching on the chemical
potential favors the pion condensate $\rho_0$, we can always set $\phi_0=0$ in
this case. The second gap equation in Eq. (\ref{renorm_gapeq_dynamic}) then
tells us that $M^2=\mu_I^2$. Plugging this into Eq.
(\ref{renorm_gapeq_constraint}), we realize that it yields an explicit
expression for the pion condensate in the chiral limit,
\begin{equation}
\rho_0^2=f_{\pi}^2-\frac{T^2}{16\pi^2}J_1(\beta\mu_I)
+\frac{\mu_I^2}{16\pi^2}\left(\frac{32\pi^2}\lambda+\ln\frac{\mu^2}{\mu_I^2}\right).
\label{LO_chilim_picond}
\end{equation}

At $\mu_I=0$, we use the fact that $J_1(0)=\frac43\pi^2$ to reduce this
expression to the simple formula \cite{Andersen:2004ae}
\begin{equation*}
\rho_0=\sqrt{f_\pi^2-\frac{T^2}{12}},
\end{equation*}
which predicts a second-order chiral-symmetry-restoring phase transition at
$T_c=\sqrt{12}f_\pi$. At nonzero chemical potential, the critical temperature
cannot be calculated analytically. However, at weak coupling it will be
presumably much larger than $\mu_I$ so that we can still approximate
$J_1(\beta\mu_I)$ with $\frac43\pi^2$. We then obtain a weak-coupling result
for the critical temperature,
\begin{equation*}
T_c^2=12\left(f_\pi^2+\frac{2\mu_I^2}\lambda\right),
\end{equation*}
which is consistent with calculations using different methods
\cite{Kapusta:1981aa,Andersen:2006ys}, and justifies the assumption we made,
$T_c\gg\mu_I$.

\subsubsection{Physical point}
While in the chiral limit the pion condensate persists down to zero temperature
and chemical potential, i.e., the vacuum, when nonzero $H$ breaks the symmetry
explicitly, the situation changes. The chiral condensate is now always nonzero,
$\phi_0^2=H/M^2$, and Eq. (\ref{LO_chilim_picond}) modifies to
\begin{equation*}
\rho_0^2=f_{\pi}^2-\frac{H^2}{\mu_I^4}-\frac{T^2}{16\pi^2}J_1(\beta\mu_I)
+\frac{\mu_I^2}{16\pi^2}\left(\frac{32\pi^2}\lambda+\ln\frac{\mu^2}{\mu_I^2}\right).
\end{equation*}
This shows that, as expected, the pion condensate increases with chemical
potential and decreases with temperature. However, at $\mu_I\to0$ the
right-hand side of this equation goes to $-\infty$. That means, there is a
critical chemical potential $\mu_{Ic}$, given by the solution of
\begin{equation}
f_{\pi}^2-\frac{H^2}{\mu_{Ic}^4}-\frac{T^2}{16\pi^2}J_1(\beta\mu_{Ic})
+\frac{\mu_{Ic}^2}{16\pi^2}\left(\frac{32\pi^2}\lambda+\ln\frac{\mu^2}{\mu_{Ic}^2}
\right)=0,
\label{critchem}
\end{equation}
below which there is no pion condensate; this is the normal phase. In that case
$M^2$ is found from the gap equation (\ref{renorm_gapeq_constraint}) by
substituting for $\phi_0$ from Eq. (\ref{renorm_gapeq_dynamic}),
\begin{equation}
f_{\pi}^2-\frac{H^2}{M^4}-\frac{T^2}{16\pi^2}J_1(\beta M)
+\frac{M^2}{16\pi^2}\left(\frac{32\pi^2}\lambda+\ln\frac{\mu^2}{M^2}\right)=0.
\label{LO_norm_sol}
\end{equation}

Note that the normal-phase solution is completely independent of the chemical
potential. Remembering that at the leading order, the parameter $M$ has the
meaning of the pion mass in the vacuum \cite{Andersen:2004ae}, the observation
that Eqs. (\ref{critchem}) and (\ref{LO_norm_sol}) are identical upon the
replacement $\mu_I\leftrightarrow M$ immediately leads to the physically
expected conclusion that the critical chemical potential for pion condensation
is equal to the pion mass in the vacuum.

\subsubsection{Numerical results}
\begin{figure}
\includegraphics[scale=0.35]{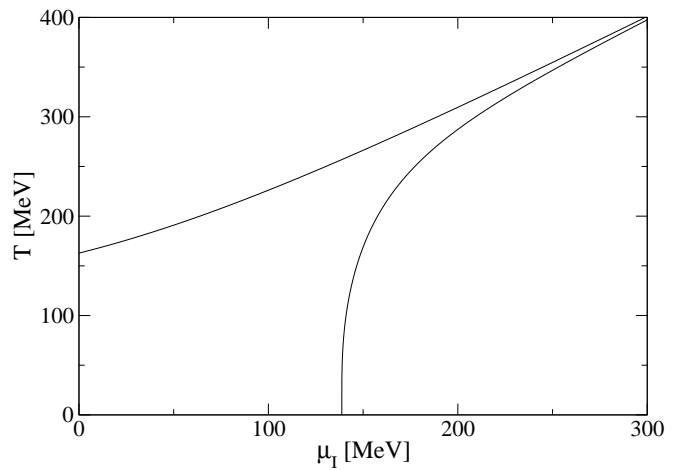}
\caption{Phase diagram for pion condensation at the leading order. The upper curve is the
chiral limit and the lower curve corresponds to the physical point.}
\label{Fig:LO_phdiag}
\end{figure}
We start with discussion of the choice of the parameters in the
Lagrangian~(\ref{l1}). First of all, in order for the theory to be stable, the
bare coupling $\lambda_b$ has to be positive. This implies that the maximum
value the cutoff $\Lambda$ may take at fixed renormalized coupling $\lambda$,
is
\begin{equation*}
\Lambda_{\text{max}}=\mu\exp\left(\frac{16\pi^2}\lambda\right).
\end{equation*}
In Refs.~\cite{Andersen:2004ae,Warringa:2006rn}, it was shown that the mass of
the sigma predicted by the model is maximal at $\lambda(\mu=100\text{
MeV})=80$, in which case $m_\sigma=433\text{ MeV}$, lying rather low in the
experimentally observed range $m_\sigma=400-800\text{ MeV}$. However, for this
large coupling the cutoff is forced to be unphysically small since
$\Lambda_{\text{max}}=720\text{ MeV}$. We therefore choose a lower value,
$\lambda(\mu=100\text{ MeV})=30$. The symmetry-breaking parameter $H$ is
adjusted to $H=(104\text{ MeV})^3$ so that it, together with the pion decay
constant $f_\pi=47\text{ MeV}$ (note that our definition of $f_\pi$ differs by
a factor of $1/2$ from the standard value), reproduces the average of the
measured masses of $\pi^0,\pi^\pm$, $m_\pi=138\text{ MeV}$.

Note that for these values of the parameters, the sigma mass comes out as
$m_\sigma=350\text{ MeV}$. The discrepancy with the measured value might be due
to the fact that we restrict our treatment to the pion sector, and therefore
miss contributions from the third-flavor physics. On the other hand, it might
as well be just a shortcoming of the present approach.

\begin{figure}
\includegraphics[scale=1.1]{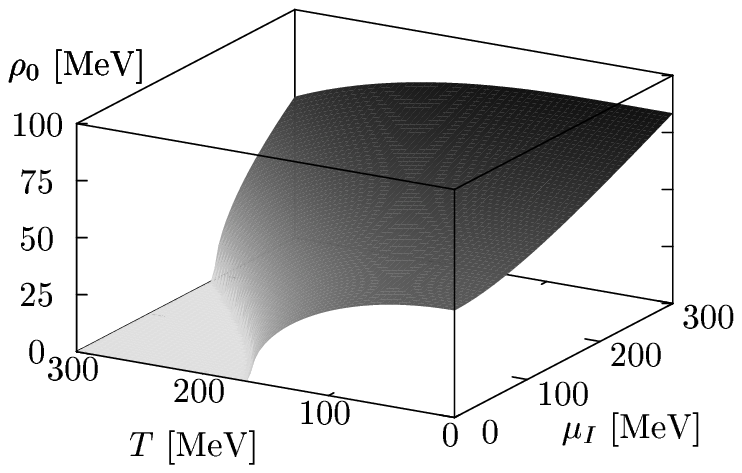}
\caption{Pion condensate in the chiral limit as a function of temperature and chemical
potential.}
\label{Fig:LO_pi_chiral}
\end{figure}

\begin{figure}
\includegraphics[scale=1.1]{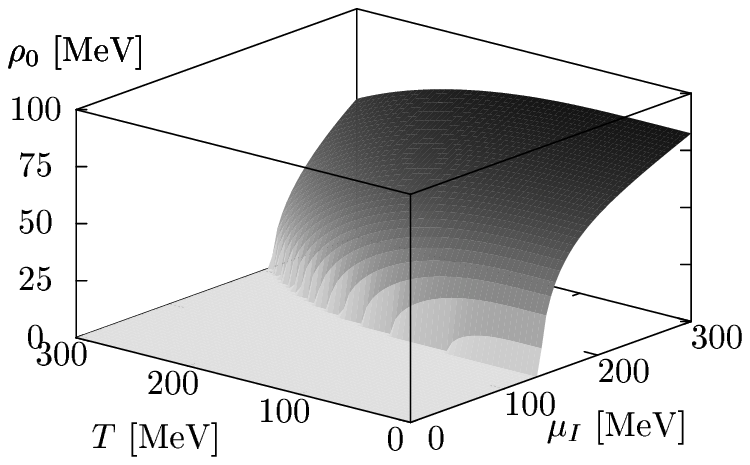}
\caption{Pion condensate at the physical point as a function of temperature and chemical
potential.}
\label{Fig:LO_pi_physical}
\end{figure}

With the above remark in mind, we plot in Fig. \ref{Fig:LO_phdiag} the LO phase
diagram both in the chiral limit and for the physical value of $H$. This phase
diagram is identical to that found previously in Ref. \cite{Andersen:2006ys}
using the 2PI effective action, because the leading orders of the 1PI and 2PI
formalisms coincide. In Figs. \ref{Fig:LO_pi_chiral} and
\ref{Fig:LO_pi_physical} we show the dependence of the pion condensate on
temperature and chemical potential, in the chiral limit and at the physical
point, respectively.

\section{Next-to-leading order}
\label{Sec:NLO}

\subsection{Effective potential}
\label{Subsec:NLO_effpot} The NLO contribution to the effective potential is
given by the second line of Eq. (\ref{matrix}). By performing the Gaussian
integral over the field $\chi$, we obtain
\begin{multline}
V_{\text{NLO}}=-\frac12\sumint_P\ln[I(P,M)]+\frac12M^2a_1+\frac12M^4b_1\\
-\frac12\sumint_P\ln\left[(P^2+m^2)^2+4\mu_I^2\omega_n^2\right]
+\sumint_P\ln(P^2+M^2).
\label{VNLO}
\end{multline}
Here $I(P,M)$ is the inverse propagator of the $\alpha$-field in momentum space
and reads
\begin{multline}
I(P,M)=\frac12\Pi(P,M)+\frac1\lambda+b_0+\frac{\phi_0^2}{P^2+M^2}\\
+\frac{\rho_0^2(P^2+m^2)}{(P^2+m^2)^2+4\mu_I^2\omega_n^2}.
\label{idef}
\end{multline}
The last two terms of Eq. (\ref{VNLO}) represent the correction to the pressure
of a free gas of charged scalar bosons, induced by finite chemical potential.
It stems from the fact that at the leading order, $\phi_1$ and $\phi_2$ were
treated on the same footing as all other fields, despite the fact that they are
endowed with the chemical potential. The finite-temperature part of this
correction (i.e., the difference of the Matsubara sum and the corresponding
frequency integral) reduces to
\begin{equation*}
\frac{T^4}{64\pi^2}\left[K_0^+(\beta M,\beta\mu_I)+K_0^-(\beta M,\beta\mu_I)
-2J_0(\beta M)\right],
\end{equation*}
where the functions $K_0^\pm(\beta M,\beta\mu_I)$ are the finite-$\mu_I$ generalizations
of $J_0(\beta M)$,
\begin{equation*}
K_0^\pm(\beta M,\beta\mu_I)=\frac{32}{3T^4}\int_0^\infty
dp\,\frac{p^4}{\omega_p}n(\omega_p\pm\mu_I).
\end{equation*}
On the other hand, the $T=0$ part of the last two terms of Eq. (\ref{VNLO})
gives, when evaluated with a four-dimensional cutoff, the contribution
\begin{equation*}
\frac1{32\pi^2}\left(\mu_I^2\Lambda^2-\mu_I^2M^2+\frac{\mu_I^4}6\right).
\end{equation*}
This is clearly unphysical since it gives, among others, a $\mu_I$-dependent
quadratic divergence. Should we use unconstrained integration over the
frequency in combination with three-dimensional cutoff or dimensional
regularization in $3-\epsilon$ space dimensions, we would get simply zero
instead of these unpleasant terms \cite{Kapusta:1989ka}. However, with these
regularization schemes, which do not bound the size of the four-momentum, we
would run into trouble with the self-energy $\Pi(P,M)$: As we will see shortly,
it would turn negative at large enough $P^2$, due to the existence of the
Landau pole in the running coupling. This clash of low-energy and high-energy
physics prevents us from applying our method directly to the thermodynamics at
nonzero chemical potential \footnote{Nevertheless, it should be stressed that
the strategy for extracting the UV-divergent part of the effective potential
would formally not change even at finite chemical potential. The only necessary
modification of the following formulas for the NLO effective potential would be
the natural replacement $\phi_0^2\to\phi_0^2+\rho_0^2$ in the definition of the
quantity $G$. The divergent part of NLO effective potential (\ref{nlodiv}), as
well as the whole NLO renormalization procedure including the prescription
(\ref{NLO_counterterms}), would remain unchanged.}. Therefore, from now on we
set $\mu_I=0$ and investigate the NLO corrections to the thermodynamics of
chiral symmetry breaking. Moreover, at zero chemical potential we can, without
lack of generality, set the pion condensate to zero, $\rho_0=0$.

The expression for the NLO effective potential (\ref{VNLO}) thus reduces just
to the first term plus the counterterms. We first evaluate the self-energy
function $\Pi(P,M)$, which is involved in the effective inverse propagator of
the $\alpha$ field. Summing over Matsubara frequencies and integrating over
three-momentum, we obtain
\begin{widetext}
\begin{equation}
\Pi(P,M)=\frac1{16\pi^2}\left(\ln\frac{\Lambda^2}{M^2}+1
+\sqrt{\frac{P^2+4M^2}{P^2}}\ln\frac{\sqrt{P^2+4M^2}-\sqrt{P^2}}
{\sqrt{P^2+4M^2}+\sqrt{P^2}}\right)+\Pi_T(P,M),
\label{divpt}
\end{equation}
where $\Pi_T(P,M)$ is the temperature-dependent part of the self-energy,
\begin{equation*}
\Pi_T(P,M)=\frac1{8\pi^2 p}\int_0^{\infty}dq\,\frac q{\omega_q}
\ln\left(\frac{q^2+pq+A^2}{q^2-pq+A^2}\right)n(\omega_q),\quad
A^2=\frac{P^4+4M^2p_0^2}{4P^2}.
\end{equation*}
\end{widetext}
The divergence in the expression (\ref{idef}) coming from the self-energy
(\ref{divpt}) is cancelled by the LO counterterm $b_0$.

Proceeding with the same steps as in Ref. \cite{Andersen:2004ae}, we next
expand the self-energy in powers of $1/P^2$ in order to extract the dominant
ultraviolet (UV) contributions. After averaging over angles, the UV-divergent
part of the NLO effective potential stems from the part of the
$\alpha$-propagator that we call $I_{\text{UV}}(P,M)$,
\begin{equation*}
\ln[I_{\text{UV}}(P,M)]=\ln C_1+\frac1{P^2}\frac{C_2}{C_1}
-\frac1{2P^4}\frac{C_2^2}{C_1^2}+\frac1{P^4}\frac{C_3}{C_1},
\end{equation*}
where
\begin{align*}
C_1=&\ln\frac{\Lambda^2}{P^2}+\kappa,\\
C_2=&2(G-2M^2)+2M^2C_1,\\
C_3=&-2M^2(G-\tfrac{M^2}2)-2M^4C_1,
\end{align*}
and we have defined
\begin{align*}
\kappa&=1+32\pi^2\left(\frac1{\lambda}+b_0\right),\\
G&=16\pi^2\phi_0^2+T^2J_1(\beta
M)-M^2\ln\frac{\mu^2}{M^2}-\frac{32\pi^2M^2}\lambda.
\end{align*}

In order to avoid infrared (IR) singularity in the $1/P^4$ term, we make the
replacement $P^4\to(P^2+\delta^2)^2$ in the denominator, where $\delta$ is a
small IR cutoff. Needless to say that this is just an artifact of the $1/P^2$
expansion and the full expression for the NLO effective potential, of course,
does not depend on $\delta$. We next integrate $\ln[I_{\text{UV}}(P,M)]$ over
$P$ and extract the UV-divergent part. The result can be written as
\begin{equation}
-\frac12\int_P\ln[I_{\text{UV}}(P,M)]=D+\text{UV-finite pieces},
\label{UVfinite}
\end{equation}
where the divergent terms are contained in
\begin{multline}
D\equiv\int_P\biggl[-\frac1{P^2}\biggl(M^2+\frac{G-2M^2}{\kappa+\ln\frac{\Lambda^2}{P^2}}
\biggr)+\frac{2M^4}{(P^2+\delta^2)^2}\biggr]\\
=\frac1{16\pi^2}\left[-M^2\Lambda^2+(G-2M^2)\Lambda^2e^\kappa\li\frac1{e^\kappa}\right.\\
\left.+2M^4\left(\ln\frac{\Lambda^2}{\delta^2}-1\right)\right].
\label{nlodiv}
\end{multline}
We have provided an exact representation of the divergence $D$, which is a
necessity for the numerical implementation of renormalization.

\subsection{Renormalization}
\label{Subsec:NLO_ren} The UV-divergent term in Eq. (\ref{nlodiv}),
proportional to $G$, depends explicitly on temperature, except at the solution
of the LO gap equation (\ref{renorm_gapeq_constraint}), where $G$ reduces to
$16\pi^2f_\pi^2$. This led previously to the conclusion that the NLO effective
potential is renormalizable in a temperature-independent manner only at the
minimum \cite{Andersen:2004ae}. At this point, we should carefully distinguish
what effective potential we are speaking of. It is crucial to note that what we
have actually calculated so far, is not an effective potential of just the
scalar fields $\phi_i$, but one of $\phi_i$ \emph{and} $\alpha$.

As is clear already from the LO expression (\ref{VLO}), such an effective
potential does not have the usual physical interpretation as the minimum energy
at fixed vacuum expectation values of the fields of the theory. In fact, it is
unbounded from both below and above and the solution of the LO gap equations
(\ref{renorm_gapeq_dynamic}), (\ref{renorm_gapeq_constraint}) is its saddle
point rather than a global extremum. This problem can be traced back to the
fact that the field $\alpha$ is not an independent dynamical degree of freedom;
upon using its equation of motion, it turns out to be a composite operator
function of $\phi_i$.

The way out is to treat the gap equation for $M$ (\ref{constraint_gapeq}) as
merely a constraint that serves to eliminate $M$ in favor of $\phi_0$. We thus
arrive at an effective potential as a function solely of $\phi_0$, which
already has the usual properties \cite{Coleman:1974jh}. This is the physical
effective potential, $\pV$, we eventually want to calculate, independent of the
method we use.

First of all, let us note that none of the results of Sec. \ref{Sec:LO} is
altered by changing the perspective from $V_{\text{LO}}(M,\phi_0)$ to
$\pV_{\text{LO}}(\phi_0)$. The reason is that, of course, the stationary point
of $\pV_{\text{LO}}$ coincides with that of $V_{\text{LO}}$ once the LO
constraint (\ref{renorm_gapeq_constraint}) has been used to eliminate $M$ from
it.

At the next-to-leading order we have to be more careful. Expanding the solution
of the constraint (\ref{constraint_gapeq}) in powers of $1/N$ (and suppressing
the argument in $M(\phi_0)$ for the sake of legibility), the physical effective
potential up to next-to-leading order reads,
\begin{widetext}
\begin{multline*}
\pV(\phi_0)=NV_{\text{LO}}(M_{\text{LO}}+\tfrac1NM_{\text{NLO}}+\dotsb,\phi_0)
+V_{\text{NLO}}(M_{\text{LO}}+\tfrac1NM_{\text{NLO}}+\dotsb,\phi_0)\\
=NV_{\text{LO}}(M_ {\text{LO}},\phi_0)+\frac{\de
V_{\text{LO}}}{\de M}(M_{\text{LO}},\phi_0)M_{\text{NLO}}+
V_{\text{NLO}}(M_{\text{LO}},\phi_0)+\mathcal{O}(1/N)\\
=NV_{\text{LO}}(M_ {\text{LO}},\phi_0)+V_{\text{NLO}}(M_{\text{LO}},\phi_0)
+\mathcal{O}(1/N),
\end{multline*}
\end{widetext}
where the middle term vanishes due to the LO constraint
(\ref{renorm_gapeq_constraint}). We can thus see that in order to calculate the
physical effective potential to next-to-leading order, it is sufficient to
solve the LO constraint for $M$. Doing that, we obtain the effective potential
as a function of $\phi_0$ alone. Moreover, Eq. (\ref{nlodiv}) tells us that the
divergences of $\pV_{\text{LO}}$ are independent of temperature for all values
of $\phi_0$ since $G=16\pi^2f_\pi^2$ follows directly from the constraint
(\ref{renorm_gapeq_constraint}) and does not need the true gap equation
(\ref{renorm_gapeq_dynamic}) to be fulfilled. Substituting for $G$, Eqs.
(\ref{nlodiv}) and (\ref{VNLO}) show that the necessary NLO counterterms are
\begin{equation}
a_1=\frac{\Lambda^2}{8\pi^2}\left(1+2e^\kappa\li\frac1{e^\kappa}\right),\quad
b_1=-\frac1{4\pi^2}\ln\frac{\Lambda^2}{\mu^2}.
\label{NLO_counterterms}
\end{equation}

Finally, it is important to check whether the procedure of elimination of $M$
is in practice well defined. To this end, note that the left-hand side of Eq.
(\ref{renorm_gapeq_constraint}) is a monotonically increasing function of
$M^2$. It goes to $+\infty$ as $M^2\to+\infty$, and reaches the limit
$f_\pi^2-\phi_0^2-\frac{T^2}{12}$ as $M^2\to0$. This means that the constraint
has a real solution only when $\phi_0$ is such that $\phi_0^2\geq
f_\pi^2-\frac{T^2}{12}$. If this condition is satisfied, the solution of the
constraint is unique as it should. Inside the interval $(-R,+R)$ around the
origin with the ``radius'' $R=\sqrt{f_\pi^2-\frac{T^2}{12}}$, the effective
potential $\pV(\phi_0)$ is undetermined by the auxiliary field technique. In
principle it can be analytically continued to this interval \cite{Sher:1988mj}
(this is what we actually do in our weak-coupling analysis in Sec.
\ref{Subsec:NLO_gapeq}), but that would be hard to implement numerically.
Fortunately, for our parameter set the minimum of the NLO effective potential
always lies outside this interval so that the problem does not arise. It is
also useful to note that the interval $(-R,+R)$ is precisely the region where
the scalar field fluctuations become tachyonic, and the perturbative one-loop
effective potential acquires nonzero imaginary part.

\subsection{Gap equations}
\label{Subsec:NLO_gapeq} Once we have renormalized the NLO effective potential,
we may proceed in the standard manner and calculate the condensate $\phi_0$ as
a solution to the gap equation,
\begin{equation}
\frac{d\pV}{d\phi_0}=0.
\label{gapeq_general}
\end{equation}
Thanks to the fact that $M$ now depends on the condensate, we have to evaluate
the total derivative using the chain rule, which leads to
\begin{equation}
\frac{\de V_{\text{LO}}}{\de\phi_0}+\frac1N\left(\frac{\de V_{\text{NLO}}}{\de\phi_0}+
\frac{\de V_{\text{NLO}}}{\de M^2}\frac{\de M_{\text{LO}}^2}{\de\phi_0}\right)=0.
\label{gapeq_physical}
\end{equation}
It is important to observe that while both $\de V_{\text{NLO}}/\de\phi_0$ and
$\de V_{\text{NLO}}/\de M^2$ may in general contain temperature-dependent
divergences (just because we can set $G=16\pi^2f_\pi^2$ only \emph{after}
taking the derivative), these cancel in the linear combination in Eq.
(\ref{gapeq_physical}).

To illustrate this procedure by an explicit example, let us consider the
weak-coupling limit in case of exact chiral symmetry, i.e., $H=0$. Since $M=0$
in the chirally broken phase at the stationary point of the LO effective
potential (\ref{VLO}), we expect it to be of order $\mathcal{O}(\lambda)$ at
the NLO value of the chiral condensate. Then, in the LO constraint
(\ref{renorm_gapeq_constraint}) we can set $M=0$ everywhere except the term
proportional to $1/\lambda$, which leads to the expression for $M$,
\begin{equation}
M^2=\frac\lambda2\left(-f_\pi^2+\phi_0^2+\frac{T^2}{12}\right).
\label{weak_coupling_M}
\end{equation}
The first partial derivative in Eq. (\ref{gapeq_physical}) is simply
$-M^2\phi_0$, while the last one is $\lambda\phi_0$ by Eq.
(\ref{weak_coupling_M}). The $\mathcal{O}(\lambda)$ part of the derivative of
$V_{\text{NLO}}$ with respect to $\phi_0$ is given by
\begin{multline*}
\frac{\de
V_{\text{NLO}}}{\de\phi_0}=-\phi_0\sumint_P\frac1{P^2+M^2}\\
\times\frac1{\frac12\Pi(P,M)+\frac1\lambda+b_0+\frac{\phi_0^2}{P^2+M^2}}
\to-\lambda\phi_0\sumint_P\frac1{P^2+M^2}.
\end{multline*}
By an analogous argument, the partial derivative with respect to $M^2$ is
$\frac12a_1+\mathcal{O}(\lambda)\to\Lambda^2/16\pi^2$ as $\lambda\to0$. Putting
all the pieces together, we find
\begin{multline*}
\frac{d\pV}{d\phi_0}=-M^2\phi_0-\frac\lambda N\phi_0\left(\sumint_P\frac1{P^2+M^2}-
\frac{\Lambda^2}{16\pi^2}\right)\\
=-\phi_0\left(M^2+\frac\lambda N\frac{T^2}{12}\right)+\mathcal{O}(\lambda^2).
\end{multline*}
The sum-integral above was calculated by differentiation of Eq.
(\ref{sumintlog}) with respect to $M^2$. This, together with the LO constraint
(\ref{weak_coupling_M}), immediately yields the NLO expression for the chiral
condensate in the weak-coupling limit,
\begin{equation*}
\phi_0^2=f_\pi^2-\left(1+\frac2N\right)\frac{T^2}{12},
\end{equation*}
and the NLO critical temperature
\begin{equation}
T_c=\sqrt{\frac{12}{1+\frac2N}}f_\pi.
\label{TcNLO}
\end{equation}

It is important to stress that while the LO critical temperature,
$T_c=\sqrt{12}f_\pi$, derived in Sec. \ref{Subsec:LO_phdiag} is a
nonperturbative result valid for all values of the coupling, the present NLO
correction is just a weak-coupling limit. In fact, while Eq. (\ref{TcNLO})
predicts a \emph{decrease} of the critical temperature with respect to the LO
value, in particular for $N=4$ by about $20\%$, our numerical analysis
discussed in Sec. \ref{Subsec:NLO_numres} shows that at $\lambda(\mu=100\text{
MeV})=30$ the critical temperature actually \emph{increases} by more than
$30\%$! In order to make sure that these two results are consistent, we will
briefly investigate the dependence of the critical temperature on the coupling.

Apart from calculating the exact minimum of the NLO effective potential, the
NLO correction to the chiral condensate can also be determined in a more
straightforward manner \cite{Veillette:2007ve}. One writes the solution to Eq.
(\ref{gapeq_general}) as $\phi_0=\phi_0^{\text{LO}}+\frac1N\phi_0^{\text{NLO}}$
and expands the whole gap equation in powers of $1/N$. Comparing the terms of
the same order, one thus gets an explicit formula for the NLO correction to the
condensate,
\begin{equation}
\phi_0^{\text{NLO}}=-\frac{\frac{d\pV_{\text{NLO}}(\phi_0^{\text{LO}})}{d\phi_0}}
{\frac{d^2\pV_{\text{LO}}(\phi_0^{\text{LO}})}{d\phi_0^2}}.
\label{NLOgap_explicit}
\end{equation}
While the first derivative of the NLO effective potential in the numerator of
Eq. (\ref{NLOgap_explicit}) is best evaluated numerically, the second
derivative of the LO effective potential in the denominator can be calculated
analytically. Keeping in mind that this is a total derivative we find, in a
fashion similar to Eq. (\ref{gapeq_physical}),
\begin{equation*}
\frac{d^2\pV_{\text{LO}}}{d\phi_0^2}=\frac{\de^2V_{\text{LO}}}{\de\phi_0^2}-
\left[\frac{\de^2V_{\text{LO}}}{\de(M^2)^2}\right]^{-1}\left(\frac{\de^2V_{\text{LO}}}
{\de\phi_0\de M^2}\right)^2.
\end{equation*}
At the intermediate step, we used the implicit function theorem to extract the
derivative $\de M^2_{\text{LO}}/\de\phi_0$ from the constraint
(\ref{constraint_gapeq}). The second partial derivatives are easily found to be
$\de^2V_{\text{LO}}/\de\phi_0^2=-M^2$, $\de^2V_{\text{LO}}/\de\phi_0\de
M^2=-\phi_0$, and
\begin{equation}
\frac{\de^2V_{\text{LO}}}{\de(M^2)^2}=\frac1{32\pi^2}\left[J_2(\beta
M)+\frac{32\pi^2}\lambda+\ln\frac{\mu^2}{M^2}-1\right],
\label{diff_wrt_alpha}
\end{equation}
where
\begin{equation*}
J_2(\beta M)=-T^2\frac{dJ_1(\beta M)}{dM^2}
=4\int_0^{\infty}dp\,\frac{n(\omega_p)}{\omega_p}.
\end{equation*}

\subsection{Numerical results}
\label{Subsec:NLO_numres} For sake of numerical computation, the NLO effective
potential was renormalized as follows. First, the ``vacuum'' effective
potential, evaluated at zero temperature and with the LO value of the chiral
condensate, was subtracted. This canceled the leading, quartic,
field-independent divergence. In addition, we subtracted the integral
representation of the remaining divergence (\ref{nlodiv}). This amounts to NLO
renormalization according to Eq. (\ref{NLO_counterterms}) as well as a finite
piece, which must be added to the effective potential after the numerical
integration,
\begin{equation*}
\frac{M^4}{8\pi^2}\left(\ln\frac{\mu^2}{\delta^2}-1\right)-
\frac{M_0^4}{8\pi^2}\left(\ln\frac{\mu^2}{\delta^2}-1\right),
\end{equation*}
where $M_0$ refers to the value of the mass parameter in the vacuum.

Note that the UV-finite pieces indicated in Eq. (\ref{UVfinite}) depend
explicitly on the counterterm $b_0$, hence also on the cutoff. It is
technically advantageous to keep these terms and regulate the integrations with
the cutoff even after all divergences have been subtracted. Physically this
means that we treat our model as a low-energy effective theory and neglect
contributions suppressed by inverse powers of the cutoff. In Ref.
\cite{Andersen:2004ae} the cutoff dependence of the final results was shown to
be mild enough to justify this procedure. In concrete calculations, we chose
$\Lambda=5\text{ GeV}$.

\begin{figure}
\includegraphics[scale=0.45]{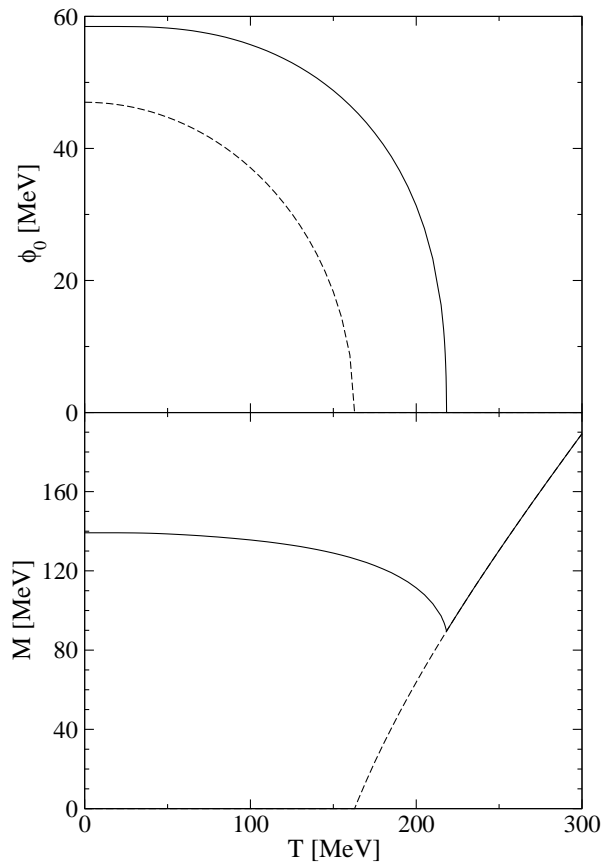}
\caption{Solution of the NLO gap equation in the chiral limit. Upper panel: the chiral
condensate; lower panel: the mass parameter $M$. For reference, the LO values are shown
by the dashed lines.}
\label{Fig:NLO_chiral}
\end{figure}

In Fig. \ref{Fig:NLO_chiral} we display the numerical solution of the NLO gap
equation (\ref{gapeq_physical}) in the chiral limit. We can see that the
correction to the chiral condensate is quite large. However, at zero
temperature, the increase is about $25\%$, which is exactly what one would
expect from a $1/N$ correction at $N=4$. The critical temperature increases
from the LO value $163\text{ MeV}$ to $218\text{ MeV}$, which means an increase
by about a third. That is still compatible with the $1/N$ prediction up to a
numerical factor close to one.

\begin{figure}
\includegraphics[scale=0.45]{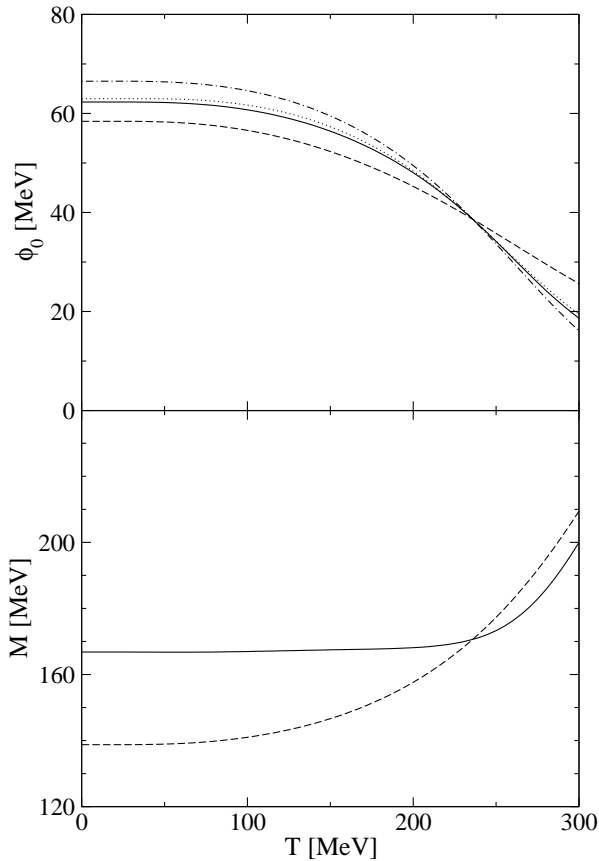}
\caption{Solution of the NLO gap equation at the physical point (solid lines). Upper
panel: the chiral condensate; lower panel: the mass parameter $M$. For reference, the LO
values are shown by the dashed lines. We also show the NLO chiral condensate calculated
using Eqs. (\ref{NLOgap_explicit}) (dash-dotted) and (\ref{NLOgap_explicit2}) (dotted),
respectively.}
\label{Fig:NLO_physical}
\end{figure}

At the physical point, the NLO corrections are considerably smaller, see Fig.
\ref{Fig:NLO_physical}. Here we also compare the exact self-consistent solution
to the NLO gap equation (solid line) with the expansion of the condensate
according to Eq. (\ref{NLOgap_explicit}) (dash-dotted line). [In the chiral
limit it cannot be used since the function $J_2(\beta M)$ diverges at the LO
solution $M_{\text{LO}}=0$.] Apparently, the formula (\ref{NLOgap_explicit}),
though exact in the large-$N$ limit, is of little use in our case: At
temperatures below about $200\text{ MeV}$, it overshoots the NLO correction to
the condensate provided by exact solution to the gap equation
(\ref{gapeq_physical}) by nearly $100\%$. In spite of that, the two evaluations
of the NLO condensate are formally consistent for their difference is of order
$1/N^2$ of the LO value.

However, the agreement with the exact solution of the NLO gap equation can be
improved by modifying Eq. (\ref{NLOgap_explicit}) to
\begin{equation}
\phi_0^{\text{NLO}}=-\frac{\frac{d\pV_{\text{NLO}}(\phi_0^{\text{LO}})}{d\phi_0}}
{\frac{d^2\pV_{\text{LO}}(\phi_0^{\text{LO}})}{d\phi_0^2}+\frac1N
\frac{d^2\pV_{\text{NLO}}(\phi_0^{\text{LO}})}{d\phi_0^2}}.
\label{NLOgap_explicit2}
\end{equation}
This expression is easily understood. The denominator contains the second
derivative of the full (LO plus NLO) effective potential, and the numerator
contains the first derivative of the same, just because
$d\pV_{\text{LO}}(\phi_0^{\text{LO}})/d\phi_0=0$. Eq. (\ref{NLOgap_explicit2})
then gives the approximate solution to the NLO gap equation by replacing the
function $d\pV/d\phi_0$ at the point $\phi_0=\phi_0^{\text{LO}}$ with a
straight line; this is just the first step of the Newton iterative algorithm
for solving nonlinear equations.

From the point of view of $1/N$ expansion, Eq. (\ref{NLOgap_explicit2}) resums
a selected series of contributions to the gap from arbitrarily high orders. Our
comparison to exact solution of the NLO gap equation demonstrates that Eq.
(\ref{NLOgap_explicit2}) (dotted line in Fig. \ref{Fig:NLO_physical}) is
superior to the simpler expression (\ref{NLOgap_explicit}) at a little extra
computational cost (just one more evaluation of the NLO effective potential
when implemented properly).

\begin{figure}
\includegraphics[scale=0.35]{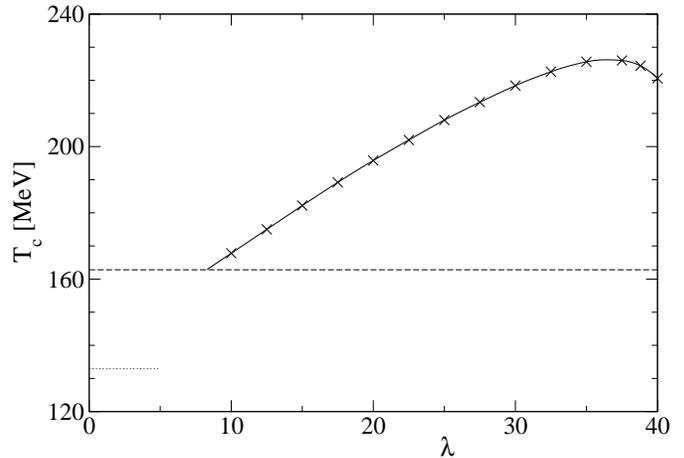}
\caption{Critical temperature for chiral symmetry breaking as a function of the coupling.
The dashed line denotes the LO critical temperature while the dotted one indicates the
weak-coupling NLO limit (\ref{TcNLO}).}
\label{Fig:Tc_lambda}
\end{figure}

In order to make connection between our numerical results and the weak-coupling
prediction (\ref{TcNLO}) for the critical temperature in the chiral limit, we
also investigated the dependence of the critical temperature on the
renormalized coupling, with all other parameters fixed. The result is shown in
Fig. \ref{Fig:Tc_lambda}. We could not check the limit (\ref{TcNLO}) directly,
because as follows from the discussion in Sec. \ref{Subsec:NLO_ren}, the
auxiliary field technique only allows us to study the range of coupling for
which the NLO critical temperature is higher than the LO one (and the chiral
condensate at fixed temperature thus larger). Nevertheless, the plot in Fig.
\ref{Fig:Tc_lambda} clearly indicates that the critical temperature decreases
at small coupling and the corresponding numerical values are consistent with
the limit (\ref{TcNLO}). (The bend at large coupling is most likely a cutoff
effect: At $\lambda=40$ the maximal cutoff allowed by stability criterion is
already very close to the actual value used in the computations.)

In addition to the nonuniversal quantities such as the critical temperature and
the condensate $\phi_0$ as a function of temperature, our NLO calculation also
allows to check some universal properties of the $\gr{O}(N)$ model. At high
temperature, the nonzero Matsubara modes decouple and the system undergoes
dimensional reduction to a Euclidean field theory in three dimensions. This
implies that the critical exponents of the system are those of the
three-dimensional $\gr{O}(N)$ model. For example, the critical exponent $\nu$
that governs the behavior of the order parameter $\phi_0$ near the critical
temperature, has been calculated in the $1/N$ expansion to next-to-leading
order \cite{Alford:2004jj,ZinnJustin:2002ru},
\begin{equation*}
\nu=\frac12-\frac4{N\pi^2}+\mathcal{O}(1/N^2).
\end{equation*}
For $N=4$, this reduces to $\nu\approx0.3987$ (as compared to the LO value
$\nu=1/2$). The power-law fit based on the NLO chiral condensate in Fig.
\ref{Fig:NLO_chiral} is in agreement with this analytical result within the
error bars.

\section{Conclusions}
\label{Sec:Concl} In the present paper, we have investigated the thermodynamics
of the $\gr{O}(N)$ linear sigma model at the next-to-leading order of the
1PI-$1/N$ expansion. We explained how the NLO effective potential can be
systematically renormalized in a temperature-independent manner. The crucial
observation leading to this result was that the effective action previously
used in literature, is one of the scalar field $\phi_i$ as well as the
auxiliary composite field $\alpha$. The latter has to be eliminated before
renormalization can be carried out consistently. One thus arrives at an
effective action which is a function solely of $\phi_i$ and has the usual
physical properties and interpretation. This procedure is further justified by
the fact that in the numerical computation, we solved the NLO gap equation by
direct extremization of the effective potential, thereby proving that the found
solution provides a thermodynamically stable configuration.

A different way to calculate the NLO corrections to the condensates, based on
expressions like (\ref{NLOgap_explicit}), is sometimes used in the literature.
This has the great advantage of being much less computationally demanding than
the self-consistent solution of the NLO gap equations, in addition to the fact
that the latter may lead to unphysical results \cite{Nikolic:2007ni}. In our
model the NLO gap equation can be solved exactly. A direct comparison of the
results obtained using the two methods shows that a careless use of Eq.
(\ref{NLOgap_explicit}) can be misleading and need not improve the
approximation provided by the leading order. We suggest a simple modification,
Eq. (\ref{NLOgap_explicit2}), which in our case yields much better results with
a little extra computational effort.

We also extended the NLO 1PI formalism to finite chemical potential. At leading
order, we determined the phase diagram and confirmed the results previously
obtained using the 2PI effective action at the leading order
\cite{Andersen:2006ys}. This is of course not surprising since at the leading
order, the 1PI and 2PI formalisms coincide. It would be interesting to compare
these two approaches at the next-to-leading order. In the present framework, we
could not extend the explicit NLO calculations to finite chemical potential due
to regularization problems. This issue as well as the relation to the 2PI
formalism \cite{Aarts:2002dj,Berges:2005hc} will be subject of future work.

\begin{acknowledgments}
J. O. A. would like to thank D. Boer and H. J. Warringa for useful discussions
and suggestions. T. B. is grateful to H. Abuki, J. Ho\v{s}ek, J. Noronha, and
D. H. Rischke for helpful discussions, and the Department of Physics at NTNU
for hospitality during two stays where part of this work was carried out. T. B.
was supported in part by the Alexander von Humboldt Foundation and by the GA CR
grant No. 202/06/0734.
\end{acknowledgments}

\end{document}